\documentclass[aps,prl,twocolumn,preprintnumbers,tightenlines,superscriptaddress]{revtex4-1}
\pdfoutput=1
\usepackage{epsfig,latexsym,cancel,amssymb,amsmath}
\usepackage{graphicx,color,hyperref}
\hypersetup{colorlinks,citecolor= nicegreen,linkcolor= nicered}
\definecolor{nicered}{rgb}{0.7,0.1,0.1}
\definecolor{nicegreen}{rgb}{0.1,0.5,0.1}

\newcommand\SEC[1]{\smallskip\noindent{\sl\bfseries #1}}

\begin{document}
\DeclareGraphicsExtensions{.jpg,.pdf,.mps,.png,}

\title{\bf Baryogenesis Once More, Inside LHC}
\author{Haipeng An}
\affiliation{Perimeter Institute, Waterloo, Ontario N2L 2Y5, Canada}
\author{Yue Zhang}
\affiliation{California Institute of Technology, Pasadena, CA 91125, USA}

\begin{abstract}
We investigate the feasibility of directly detecting a generation mechanism of the cosmic baryon asymmetry, 
by repeating the same particle physics process inside the LHC. We propose a framework with $R$-parity and CP 
violating squark decays responsible for baryogenesis, which can be embedded in supersymmetric models and 
is partly motivated by naturalness. We argue that the baryon number generation here is closely related to lepton 
charge asymmetry on the resonance. We emphasize the importance of the single charged lepton plus multijet 
channel in the absence of significant missing energy in search of such a scenario.
\end{abstract}

\preprint{CALT 68-2863}

\maketitle

\SEC{Introduction. }
The existence of baryon asymmetry in our universe is one of the mysteries and 
long-standing topics in particle physics and cosmology. Many theories have been proposed
ever since to explain the generation of the baryon asymmetry. One challenging issue 
of these mechanisms is how to detect them today. There can be various indirect 
connections to other phenomena in low energy and collider experiments, from the CP violations behind baryogenesis to the new states which lead to out-of-equilibrium conditions. While it seems difficult to reach the temperature for 
baryogenesis to happen in the current laboratories, the motivation of this work is to examine
how the particle physics side of certain baryogenesis scenarios could be detected in a more direct way.

Natural supersymmetry (SUSY) was invented to solve the hierarchy problem, 
which necessarily contains light top squarks (stop) accessible to the LHC energy. The LHC data
already constraints the stop to be heavier than $\sim$\,700 GeV~\cite{LHCstop} if it decays into neutralino, unless
one resorts to a compressed spectrum. $R$-parity violation (RPV) is an alternative option to 
hide the light stops from the current Large Hadron Collider (LHC) data~\cite{Brust:2011tb}. 
The RPV interactions controlling the lifetime of the stop may also reach equilibrium and washout the 
primordial baryon number asymmetry in the early universe~\cite{Barbier:2004ez}.
Interestingly, the minimal value of such coupling to wash out the baryon number 
coincides with the one allowing a sub-TeV stop to decay promptly at colliders~\cite{Graham:2012th}.
Therefore, the stop RPV interactions either wash out the primordial baryon asymmetry, 
or induce detectable displaced vertices in the LHC. In the first case, a new mechanism for late baryon number generation is in need. 


In this {\em Letter}, we propose a scenario for baryogenesis in which both the CP and baryon number violations can be observed at colliders, and can be realized in the natural SUSY framework. 
The RPV decay of the lightest SUSY partner (LSP) is used to generate the baryon number.
Two types of RPV interactions are needed, which break both baryon and lepton symmetries explicitly.
The LSP in charge of the genesis can be stop itself or another lighter squark, 
and it had better be lighter than a few hundred GeV for the sake of naturalness. 
This allows the LSP squark to be copiously produced at the 14 TeV LHC.
We show that successful baryogenesis requires large CP violation,
which can be manifested at collider via the lepton charge asymmetry in the decay products from the squark resonances, 
and can serve as a smoking-gun signature of this scenario.

\SEC{Toy Model. }
To capture the essence, we start with a toy model with two squarks ($i=1,2$),
\begin{eqnarray}\label{rpv0}
\mathcal{L} = \lambda''_i \bar b^c P_R c \tilde d_i + \lambda'_{ij} (\bar u_j P_R \mu^c - V_{jk} \bar d_k P_R \nu^c) \tilde d_i \ , 
\end{eqnarray}
where $j,k=1,2,3$ and  $V$ is the Cabbibo-Kobayashi-Mskawa (CKM) matrix. The above Lagrangian can be obtained from simplified version of the minimal supersymmetric SM (MSSM) with RPV, $\tilde d_i$ identified as right-handed down-type squark, and all other superpartners decoupled. 
The quark flavors in the $\lambda''$ term and the muon flavor in $\lambda'$ are chosen for illustration.
To suppress proton decay, we forbid the operators containing the first generation quarks explicitly, which implies $\lambda'_{i1}=0$ and $\lambda'_{i2} V_{21} + \lambda'_{i3} V_{31}=0$. The hierarchy of CKM element predicts $\tilde d_i$ must couple preferably to the third generation quarks, and (\ref{rpv0}) simplifies to
\begin{eqnarray}\label{rpv1}
\mathcal{L} \simeq \lambda''_i \bar b^c P_R c \tilde d_i + \lambda'_{i} (\bar t P_R \mu^c - b P_R \nu^c) \tilde d_i \ , 
\end{eqnarray}
The first thing to notice is that the existence of both $\lambda'$ and $\lambda''$ type RPV interactions induces proton decay~\cite{Smirnov:1996bg}. 
With the above choice of flavors, proton decay happens at two-loop order.
We calculate its rate by using the chiral effective Lagrangian~\cite{Claudson:1981gh} 
and the lattice results on nucleon-pion matrix elements~\cite{Aoki:1999tw}.
The leading decay mode is $p\to K$, whose partial lifetime is constrained by Super-Kamiokande~\cite{Kobayashi:2005pe}.
This translates to the upper bounds $\sqrt{|\lambda''_{i} \lambda'_i|} \lesssim 2 \times 10^{-6} \left({m_{\tilde d_i}}/{600\,\rm GeV} \right)^2$.

%

\SEC{Baryogenesis in the Early Universe. }
We proceed to discuss how baryon asymmetry could be regenerated from the decays of $\tilde d_i$ via (\ref{rpv1})~\cite{Dimopoulos:1987rk}.
The mechanism discussed here relies on $R$-parity and CP violating decays, which is similar to leptogenesis~\cite{Fukugita:1986hr, Blanchet:2009bu}, except that the decaying particles are colored and not self-conjugate, and the baryon number is created directly from their decays.
From Eq.~(\ref{rpv1}), the squarks have the decay channels
\begin{eqnarray}
\tilde d_i \to \bar b \bar c, \ t \mu^- (b \nu) \ , \ \ \ \tilde d_i^* \to b c, \ \bar t \mu^+ (\bar b \bar \nu)  \ . \label{5}
\end{eqnarray}
The $\tilde d_i$ decays generate only $B+L$, which is the difference of quantum numbers between the two final states~\cite{Nanopoulos:1979gx}. 
In order for baryon number to survive, the decay must happen after the weak sphaleron process ceases.

We define the CP violation in $\tilde d_i$ and $\tilde d_i^*$ decays and a hadronic branching ratio ${\rm Br}_{\tilde d_i \to \bar b \bar c}$ as follows
\begin{eqnarray}
\varepsilon_i \equiv \frac{\Gamma_{\tilde d_i \to \bar b \bar c} - \Gamma_{\tilde d_i^* \to b c}}{\Gamma_{\tilde d_i \to \bar b \bar c} + \Gamma_{\tilde d_i^* \to b c}} \ , \ \
{\rm Br}_i \equiv \frac{\Gamma_{\tilde d_i \to \bar b \bar c}}{\Gamma_{\tilde d_i \to \bar b \bar c} + 2 \Gamma_{\tilde d_i \to t \mu^-}}.
\end{eqnarray}
All the other decay branching ratios can be obtained from these two quantities,
%
\begin{eqnarray}\label{brs}
{\rm Br}_{\tilde d_i \to t \mu^-} \!\!&=&\! {\rm Br}_{\tilde d_i \to b \nu} \!=\! \frac {1}{2} (1 - {\rm Br}_i), \ \  
{\rm Br}_{\tilde d_i^* \to b c} \!=\! \frac{1- \varepsilon_i}{1+ \varepsilon_i} {\rm Br}_i, \nonumber \\
{\rm Br}_{\tilde d_i^* \to \bar t \mu^+} \!\!&=&\! {\rm Br}_{\tilde d_i^* \to \bar b \bar\nu} = \frac{1}{2}\left( 1- \frac{1- \varepsilon_i}{1+ \varepsilon_i} {\rm Br}_i \right) \ .
\end{eqnarray}
To generate the correct sign of baryon asymmetry requires $\varepsilon_i<0$. 
The source term of the Boltzmann equation depends on the quantity
\begin{eqnarray}
\varepsilon_i {\rm Br}_i = \frac{{\rm Im} \left(\lambda''_{i} \lambda'_{i} \lambda''^*_{j} \lambda'^*_{j} \right)}{(|\lambda''_{i}|^2 \!+\! |\lambda'_{i}|^2) (|\lambda''_{j}|^2 \!+\! |\lambda'_{j}|^2)} F_j\!\left({m_{j}^2}/{m_{i}^2}\right) ,
\end{eqnarray}
where $F_j(x) = (2\Gamma_j/m_j)[{1}/({1-x}) - 3 + (2+3x) \ln({1+1/x})]$ and $m_{i}$, $\Gamma_i$ are the mass and width of $\tilde d_i$, respectively.
When $\tilde d_1$ and $\tilde d_2$ become quasi degenerate, the resonant propagator is regularized by $\Gamma_2$~\cite{Pilaftsis:2003gt}
\begin{eqnarray}\label{Fx}
F_j(x) \approx \frac{({m_1-m_2}) (\Gamma_2/2)}{{(m_1-m_2)^2+(\Gamma_2/2)^2}} \ .
\end{eqnarray}

The Boltzmann equation to generate the baryon number is
\begin{eqnarray}\label{BoltzB}
\hspace{-0.5cm}\frac{d Y_B}{d z} \!=\! - \frac{2 \varepsilon_i \Gamma_i''}{H z} ( Y_{\tilde d_i} \!-\! Y_{\tilde d_i}^{\rm eq} ) 
\!-\! \frac{ (4 \Gamma_i'' \!+\! \Gamma_i')Y_{\tilde d_i}^{\rm eq}}{H(z) z} \frac{Y_B}{Y_q^{\rm eq}} + \cdots
%
\end{eqnarray}
where $z=M_{\tilde d_1}/T$, $Y_i \equiv n_i/s$ is the yield of $\tilde d_i$, $s = 2\pi^2 g_{*S} T^3/45$ is the total entropy density.
The term proportional to $\varepsilon_i$ is the source term, which implies $\tilde d_i$ must decay out of equilibrium.
In the washout terms, we have defined 
$\Gamma'_i \equiv 2\langle \Gamma_i  \rangle ( {\rm Br}_{\tilde d_i \to t \mu^-} + {\rm Br}_{\tilde d_i^* \to \bar t \mu^+} )$
and $\Gamma''_i \equiv \langle \Gamma_i \rangle ( {\rm Br}_{\tilde d_i \to \bar b \bar c} + {\rm Br}_{\tilde d_i^* \to b c} )$,
where $\langle \rangle$ means thermal average.
The $\cdots$ are the washout terms involving the lepton asymmetry and are numerically insignificant.

\begin{figure}[t]
\vspace{-0.3cm}
\centerline{\includegraphics[width=1\columnwidth]{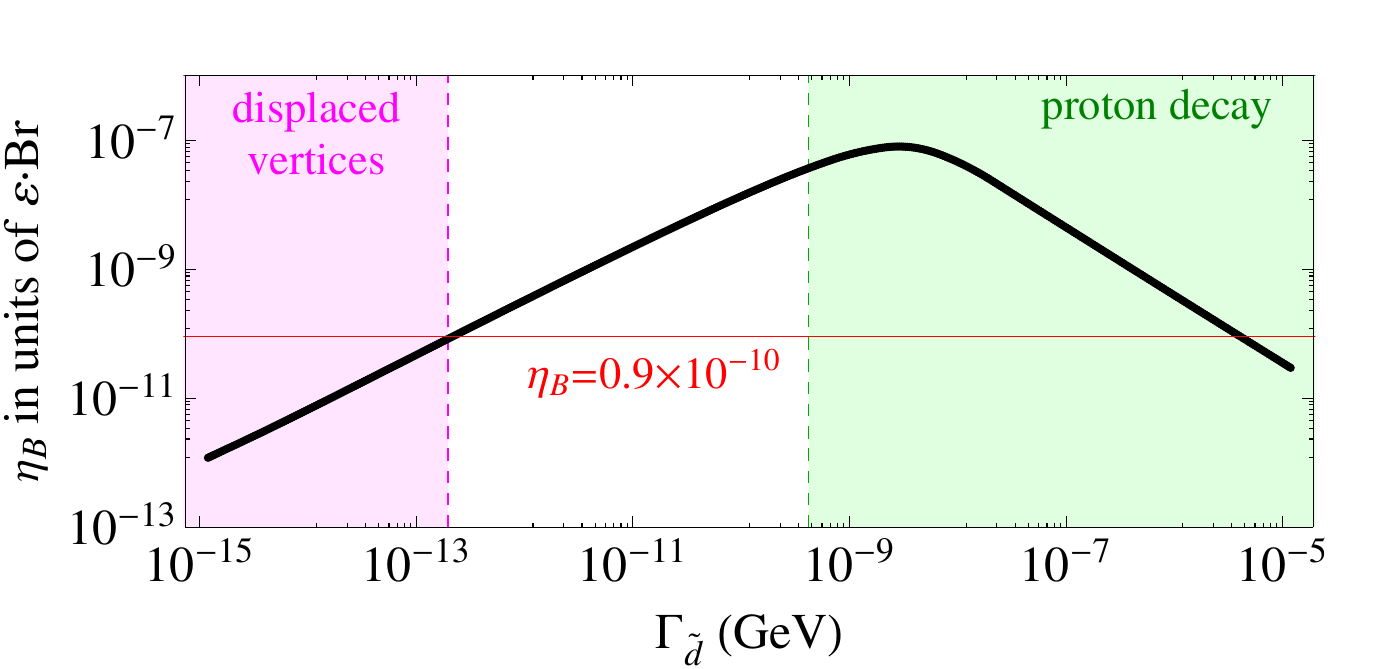}}
\vspace{-0.2cm}
\caption{Thick solid curve represents final baryon asymmetry in units of $\varepsilon \, {\rm Br}$ as a function of sdown total width, for $M_{\tilde d} =600\,$GeV.
The horizontal line is the observed baryon asymmetry.
The green region is excluded by proton decay. In the magenta region, sdown decay is displaced at LHC.}\label{CPV}
\vspace{-0.2cm}
\end{figure}

\begin{figure}[t]
\centerline{\includegraphics[width=0.55\columnwidth]{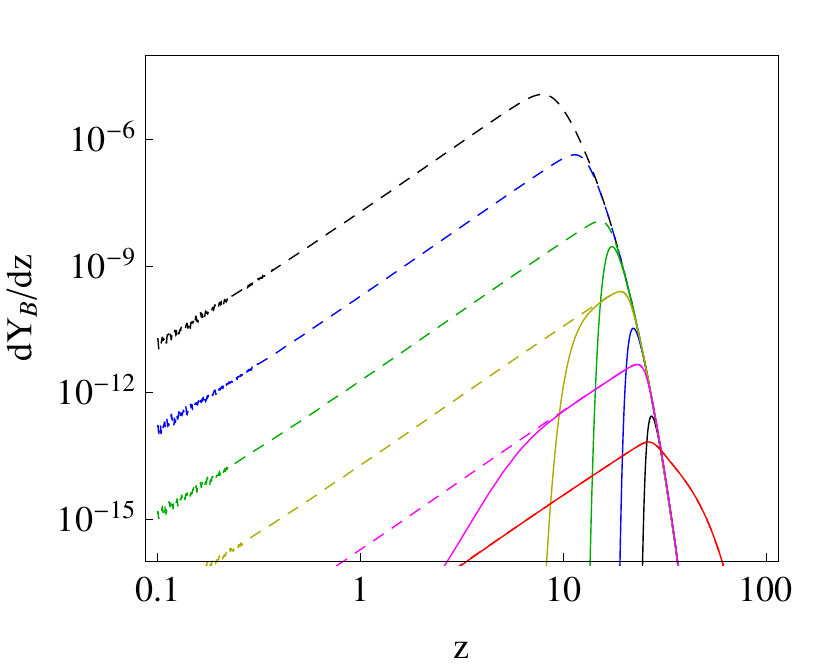}\hspace{-6mm}
\includegraphics[width=0.55\columnwidth]{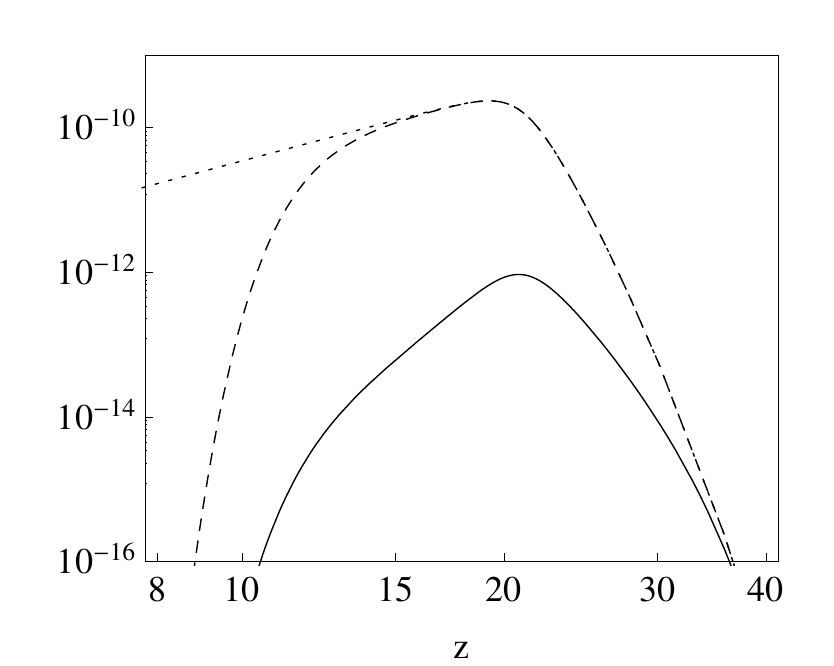}}\label{VaryLambda}
\vspace{-0.2cm}
\caption{{\bf Left:} Time (z) dependence of $d Y_B/dz$ with $m_{\tilde d}=600\,$GeV and constant $\varepsilon=1$ (solid curves). The squark width is $\Gamma_{\tilde d} = 10^{-15}$ (red), $10^{-13}$ (magenta), $10^{-11}$ (yellow), $10^{-9}$ (green), $10^{-7}$ (blue), $10^{-5}$ (black) GeV. 
The source term of Boltzmann is plotted in dashed curves. {\bf Right:} solid curve is $d Y_B/dz$ with temperature dependent $\varepsilon$ [due to (\ref{dm})], with $\Gamma_{\tilde d}=10^{-11}$\,GeV and $\Delta m_0=10^{-12}\,$GeV.}\label{VaryLambda}
\vspace{-0.2cm}
\end{figure}

Neglecting the asymmetry between the numbers of $\tilde d_i$ and $\tilde d_i^*$, which is already exponentially suppressed during the decay, the Boltzmann equation governing the $\tilde d_i$ number density can be written as
\begin{eqnarray}\label{simplifiedY}
\hspace{-0.6cm}\frac{d Y_{\tilde d_i}}{d z} = - \frac{\langle \Gamma_{i} \rangle}{H(z) z} ( Y_{\tilde d_i} - Y_{\tilde d_i}^{\rm eq}) - \frac{s \langle \sigma v_{i} \rangle}{H(z) z} ( Y_{\tilde d_i}^2 - (Y_{\tilde d_i}^{\rm eq})^2), 
\end{eqnarray}
where $\sigma v_{i}$ include all possible $\tilde d_i \tilde d_i^* \to \bar q q, g g$ annihilation channels.
The strong interaction keeps $\tilde d_i$ in equilibrium and tends to suppress $Y_{\tilde d_i} - Y_{\tilde d_i}^{\rm eq}$, and consequently the final baryon asymmetry. 
If the decay mainly happens well after the freeze-out time $z^{\rm fo}$, defined by $n_{\tilde d_i}^{\rm eq} \langle \sigma v \rangle(z^{\rm fo}) \sim H(z^{\rm fo})$, the resulting $Y_B(\infty) \sim \varepsilon Y_{\tilde d_i}^{\rm eq}(z^{\rm fo})$ will be much smaller than the observed value.

On the other hand, if the decay happens during the freeze-out, the final baryon number can be enhanced~\cite{Hambye:2012fh}. 
Combining Eqs.~(\ref{BoltzB}) and (\ref{simplifiedY}) and neglecting the washout terms,
\begin{eqnarray}
\frac{d Y_B}{d z} \sim \frac{\varepsilon_i {\rm Br}_i \langle \Gamma_i \rangle}{\langle \Gamma_i \rangle + 2 n_{\tilde d_i}^{\rm eq} \langle \sigma v_i \rangle} \frac{d Y_{\tilde d_i}^{\rm eq}}{d z} \ .
\end{eqnarray}
For $z\ll 1$, the annihilation rate dominates over the decay, 
whereas at $z\gg z^{\rm fo}$, $dY_{\tilde d_j}^{\rm eq}/dz$ is Boltzmann suppressed. Therefore, the dominant contribution to $Y_B$ is from the epoch $z\sim z^{\rm eq}$, with $n_{\tilde d_i}^{\rm eq} \langle \sigma v_i \rangle \sim \langle \Gamma_{\tilde d_i} \rangle$. The resulting baryon number is $Y_B(\infty) \sim \varepsilon_i Y_{\tilde d_i}^{\rm eq}(z^{\rm eq})$.
Thus, it can be enhanced by orders of magnitude if $z^{\rm eq}\ll z^{\rm fo}$. As a result, for a given $\varepsilon$, the final baryon asymmetry first increases with the decay rate (Fig.~\ref{CPV}), and then drops at larger decay rate because the washout terms in (\ref{BoltzB}) become important. This trend is also shown in Fig.~\ref{VaryLambda}. 

From Fig.~\ref{CPV}, we also find that successful baryogenesis requires $\varepsilon\gtrsim10^{-2}$, 
which implies the mass gap between $\tilde d_1$ and $\tilde d_2$ should be at most $10^2$ of their widths.

\SEC{Baryogenesis inside the LHC. }
As colored particles, $\tilde d_i / \tilde d_i^*$ can be copiously pair-produced at high-energy colliders. This offers a unique opportunity to access the particle physics part of the above baryogenesis scenario. We sketch the strategy of measuring the CP and lepton/baryon number violating signals at the LHC.

\begin{figure}[t!]
\centerline{\includegraphics[width=1\columnwidth]{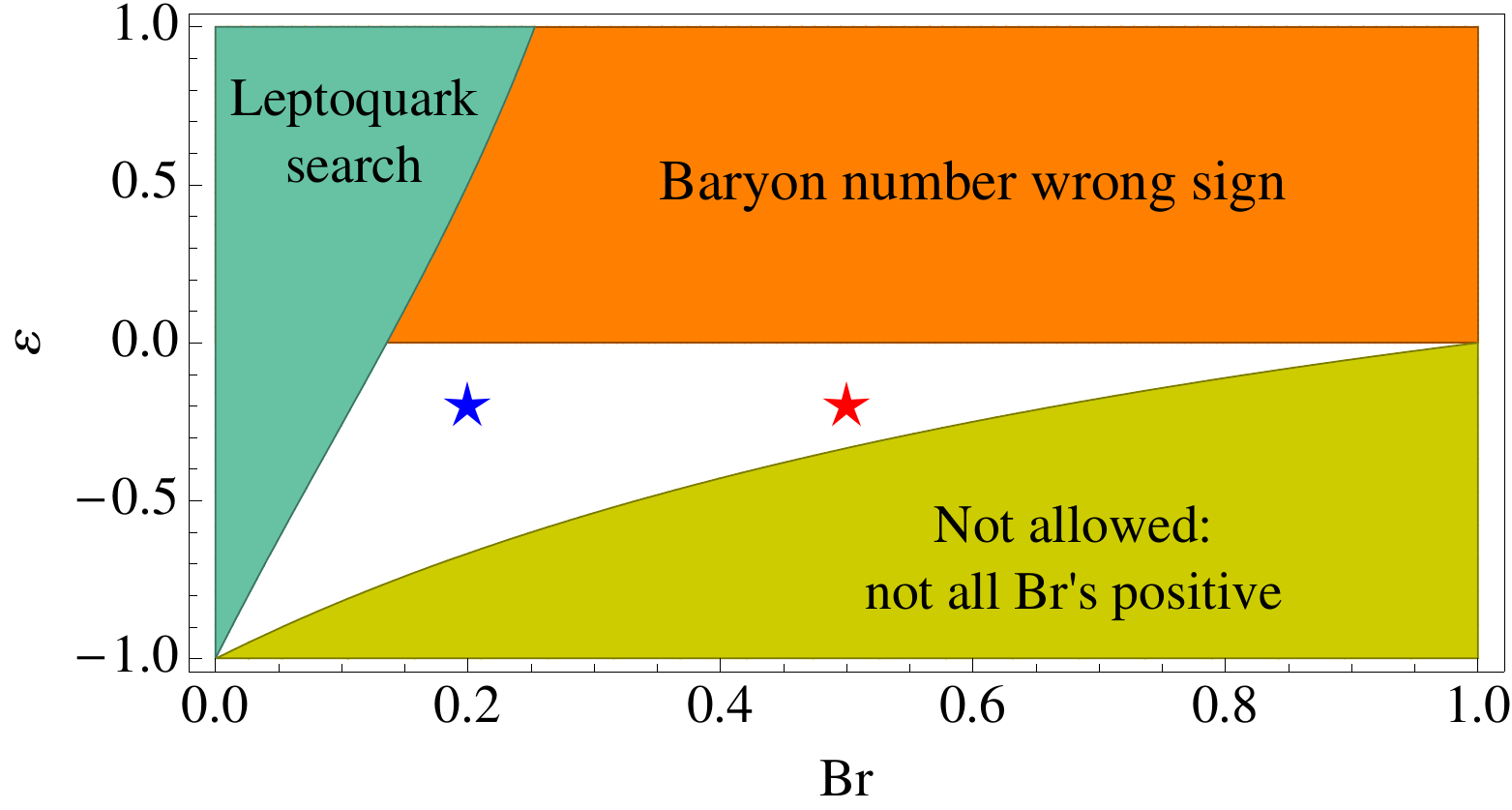}}
\vspace{-0.2cm}
\caption{Constraints on the {\rm Br}-$\varepsilon$ parameter space. The squark mass is taken equal to 600 GeV. 
The green region is excluded by the leptoquark data.
The orange region is excluded for generating a wrong sign of baryon asymmetry.
The yellow region is theoretically not accessible.}\label{all}
\vspace{-0.3cm}
\end{figure}

\smallskip
\noindent{\it Constraints. }
After being produced inside the LHC, $\tilde d_i$ and $\tilde d_i^*$ will decay according to~(\ref{5}), and the possible final states are
\begin{eqnarray}\label{fs}
\begin{tabular}{|c|c|c|}
\hline
process & signal & relevant data \\ 
\hline
$(\bar b \bar c) (b c)$ & $4j$ & ---  \\
\hline 
\hline
 & $\mu^+\mu^-2b4j$ & Leptoquark~\cite{CMSleptoquark}  \\
$(t \mu^-) (\bar t \mu^+)$ & $\mu^+\mu^-\ell^\pm2b2j\!\!\not\!\!E_T$ & Chargino- \\
 & $\mu^+\mu^-\ell^+\ell'^-2b\!\!\not\!\!E_T$ & \raisebox{0.8ex}[0pt]{Neutralino~\cite{ATLASchargino}} \\
\hline
 & $\mu^\pm2b2j\!\!\not\!\!E_T$ & Leptoquark~\cite{CMSleptoquark} \\
\raisebox{2.0ex}[0pt]{$(t \mu^-) (\bar b \bar \nu)$, $(\bar t \mu^+) (b \nu)$} & $\mu^\pm\ell^\mp2b\!\!\not\!\!E_T$ & Stop~\cite{ATLASstop} \\
\hline
$(b \nu) (\bar b \bar \nu)$, & $2b\!\!\not\!\!E_T$ & Sbottom~\cite{ATLASsbottom} \\
\hline
\hline
$(b \nu) (bc)$, $(\bar b \bar \nu)(\bar b \bar c)$ & $2b1j\!\!\not\!\!E_T$ & Multijet+$\not\!\!E_T$~\cite{CMSmultijet} \\ 
\hline
 & $\mu^\pm2b3j$ &  \raisebox{0.0ex}[0pt]{Our signal}\\
\raisebox{2.0ex}[0pt]{$(t \mu^-) (b c)$, $(\bar t \mu^+) (\bar b \bar c)$} & $\mu^\pm\ell^\mp2b1j\!\!\not\!\!E_T$ & \\
\hline
\end{tabular}\nonumber
\end{eqnarray}
with $\ell, \ell'=e, \mu$, and their branching ratios can be calculated from Eq.~(\ref{brs}). 
The corresponding LHC data relevant for the constraints are also shown.
There is no constraint from the $4j$ channel, which is the usual place to hide SUSY using RPV.
We find that the $\mu^+ \mu^- +$\,jets and $\mu^\pm+$\,jets\,$+\!\not\!\!E_T$ channels can be constrained by simple leptoquark searches~\cite{CMSleptoquark}, which give the strongest limits. 
They have been interpreted to the bounds on Br and $\varepsilon$, as displayed in Fig.~\ref{all}.

\smallskip
\noindent{\it Resonances and lepton charge asymmetry. }
Compared to other channels the analysis of $\mu^+ {\rm jets}, \ \mu^- {\rm jets}$ without significant missing energy seem to have received less motivations.
However, these channels can be used as a smoking-gun signal for the baryogenesis scenario discussed above, since $\tilde d$ decays preferably to semi-leptonic channels, whereas $\tilde d^*$ to hadronic channels. 
Therefore, if the events are triggered with a single muon and multiple hard jets, it is expected to see more $\mu^-$ events than $\mu^+$ events. 
In practice, we use {\tt PYTHIA 8}~\cite{pythia8} and {\tt FastJet 3}~\cite{fastjet3} to generate the decay events of the pair-produced $\tilde d$ and $\tilde d^*$. 
We require the transverse momenta ($P_T$) of the $\mu^\pm$ to be larger than 170 GeV.
The two hardest jets are required to have $P_T>200$ GeV, and the third hardest jet $P_T>150$ GeV.
We calculate the invariant mass of two of the jets $M_{jj}$, and compare it with the invariant mass of the muon and the rest jets $M_{\mu j}$.
By finding the combination with the closest $M_{jj}$ and $M_{\mu j}$, we identify the mass of $\tilde d$ with $M_{jj}$. 
To reduce the background from $W$+jets, we further require that the missing energy to be smaller than 30\,GeV. 
With these cuts, the major background comes from QCD multi-jet processes with one jet mis-identified as an muon. 
For $P_T > 100$ GeV, the fake rate is less than $10^{-4}$~\cite{Barr:2009zz}. To be conservative, we take the fake rate equal to this upper limit. 

\begin{figure}[t]
\vspace{-0.3cm}
\centerline{\includegraphics[width=1\columnwidth]{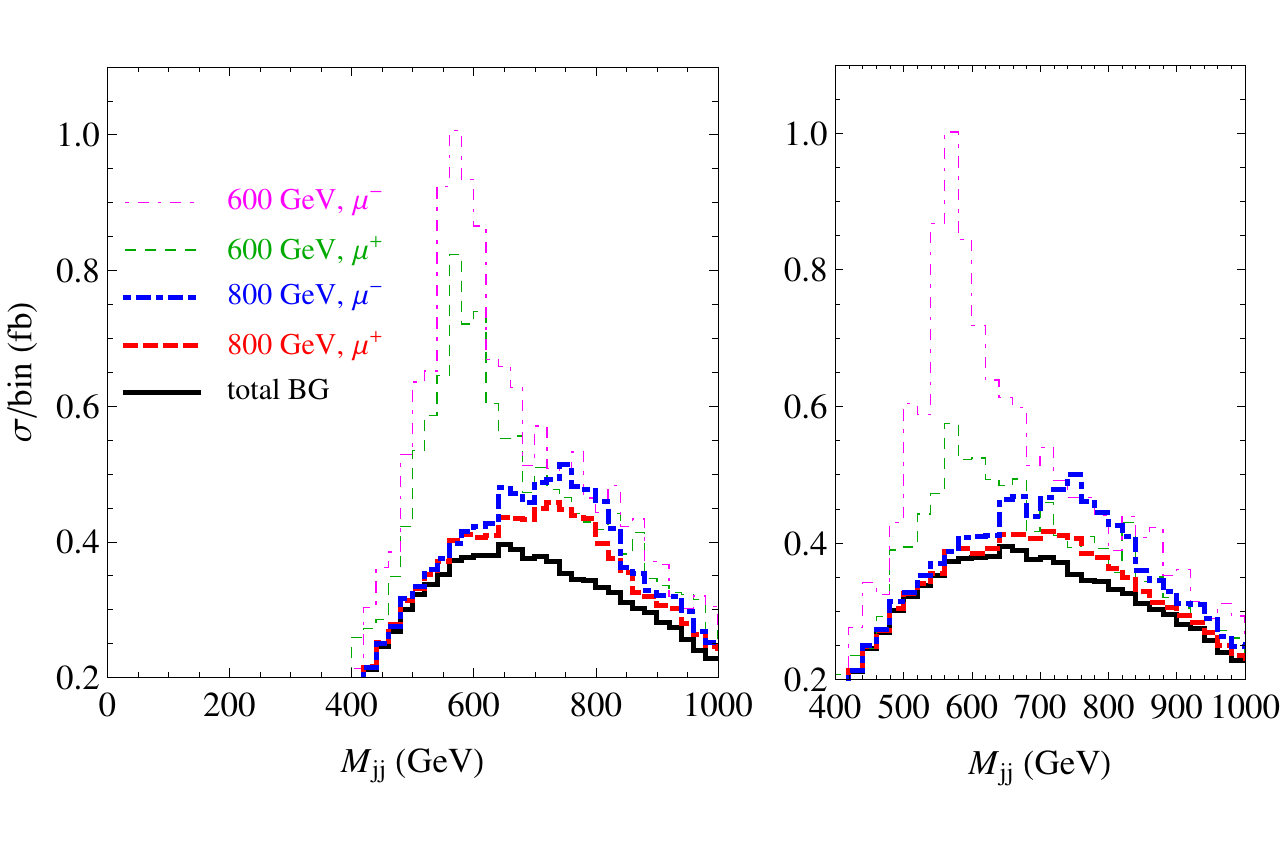}}
\vspace{-0.7cm}
\caption{Invariant mass distribution for dijet $M_{jj}$  (selected equal to $M_{\mu^\pm t}$, see text) in LHC with 14 TeV center-of-mass energy, $m_{\tilde d} =$ 600 and 800 GeV, and $\varepsilon=-0.2$, $\rm Br=0.2$ (left) and $\varepsilon=-0.2$, $\rm Br=0.5$ (right). The black curve shows the total background assuming the muon fake rate to be $10^{-4}$. Signal and background are stacked together.}\label{fig:collider}
\vspace{-0.3cm}
\end{figure}

The $M_{jj}$ distributions for events with a single $\mu^\pm$ for $m_{\tilde d}=$ 600 and 800 GeV are shown in Fig.~\ref{fig:collider}, 
where we choose two benchmark points corresponding to the blue and red stars shown in Fig.~\ref{all}. 
For the first benchmark, with $\varepsilon=-0.2$ and ${\rm Br}=0.2$, the ratio of the parton level production rates of 
the $\mu^-$\,jets and $\mu^+$\,jets events can be calculated from Eq.~(\ref{brs}) that $\hat\sigma_{\mu^-{\rm jets}}/\hat\sigma_{\mu^+{\rm jets}} \approx 1.7$. 
For the second benchmark, we have $\hat\sigma_{\mu^-{\rm jets}}/\hat\sigma_{\mu^+{\rm jets}} = 3$. 
In practice, the ratios of heights of the $\mu^-$ peak to $\mu^+$ peak shown in Fig.~\ref{fig:collider} are smaller than their parton level values, 
due to the self-contaminations from other decay channels of the resonances, apart from the SM background.
One such contamination is from the $4j$ channel, which is potentially more important, for it has the same bump structure as the desired $\mu^\pm$\,jets signal. 


\smallskip
\noindent{\it Identify baryon number generation in collider. }
From the above discussion, we are able to observe a particle or its antiparticle (on resonance) decaying into one muon and one top quark, or two hard jets. Based on the fact that a hard jet can be either a quark/antiquark or a gluon (we {\it assume} that jet substructure analysis can distinguish jets from boosted heavy particles), we enumerate all the possibilities of the color and spin quantum numbers, which can be reconstructed from the two types of final states, using $\mu(1,1/2)$, $t(3,1/2)$, $j=g(8,1)$ or $q (3,1/2)$ or $\bar q (\bar 3, 1/2)$.
The only possibilities for the quantum numbers of $(\mu t)$ and $(jj)$ to match are color triplets with integer spins. 
Meanwhile, the dijet final states decayed from the resonances are also fixed to be $(\bar q \bar q')$, $(qq')$. 
We further {\it assume} that the resonance is made of particle-anti-particle pairs ($X$ and $\bar X$).
Then, from the lepton charge asymmetry between $(\mu^-t)$ and $(\mu^+\bar t)$, we know the semi-leptonic decay branching ratios of $X$ and $\bar X$ are different,
and so are the hadronic decays due to the CPT theorem which dictates that the equality of the total widths of $X$ and $\bar X$. Therefore, a net baryon number must have been generated, if $X$ and $\bar X$ are pair produced.

Another {\it caveat} to this argument is the possibility of same sign pair productions of $XX$ and $\bar X\bar X$, with different rates, if $X$ is the superpartner of a light quark.
Because LHC is a proton-proton machine, the asymmetry of the light quark parton distributions results in the different production rates of $XX$, $\bar X\bar X$, with the exchange of a $t$-channel gluino. However, these processes are always accompanied by events with same-sign di-muon plus jets. The identification of baryogenesis at LHC requires the absence of this kind of events.


\SEC{A complete model. }
To generate enough baryon asymmetry, the mass gap between $\tilde d_1$ and $\tilde d_2$ should be of the same order as their decay widths, $\sim$\,$10^{-12}\,$GeV, which is suppressed by their RPV couplings (see Fig.~\ref{CPV}). In a natural model, such a small mass gap should be controlled by the same RPV couplings as well, which implies the existence of an approximate $SU(2)_h$ horizontal symmetry. Here we give a SUSY model example to illustrate the idea. For the sake of gauge unification, the new states should form complete representations under $SU(5)$. We add two pairs of ${\bf 5}'_i$ and ${\bf \bar 5}'_i$ to the MSSM, with $i=1,2$ forming doublets under the $SU(2)_h$ symmetry. 
We assume the $SU(2)_h$ symmetry is broken only by the RPV interactions in the superpotential
\begin{eqnarray}
\hspace{-0.4cm}W \!=\! W^{\rm RPV}_{\rm MSSM} \!+\! \lambda_i' Q_2 L_1  D'^c_i \!+\! \lambda''_i U^c_2 D^c_3 D'^c_i + M D'_i D'^c_i.
\end{eqnarray}
where $D'^c_i \in {\bf \bar 5}'_i$, and the above $\tilde d_i^*$ in the sdown case can be identified as the scalar part of $D'^c_i = \tilde d_i^* + \theta d'_i$. 
The loop generated mass difference between $\tilde d_1$ and $\tilde d_2$ is just on same order of magnitude as their widths. 
The masses of $\tilde d_i^*$ and its fermionic partner $d'_i$ can be different due to a SUSY breaking but $SU(2)_h$ conserving soft mass.
By arranging the spectrum, it is possible to allow $d'_i$ to decay into a quark and a squark via the same RPV interactions.
We find the gauge couplings are still perturbative at the unified scale in this model.

\SEC{Realization in the MSSM. }
A more interesting question is how to realize the toy model in the MSSM. Again, we need a tiny mass gap between two lightest squarks. 
This degeneracy suffers from corrections from different Yukawa couplings for different flavors, and merely requires tuning at zero temperature. 
The real problem is that, in the early universe, the finite temperature effect modifies the mass gap dynamically. 
The most important contribution comes from the F-terms of the Yukawa couplings, with the Higgs boson running in the thermal loop, 
\begin{eqnarray}\label{dm}
\hspace{-0.6cm}(m_{1} - m_{2})(T) \!&\approx&\! \Delta m_0 \!+\! \frac{y_1^2 - y_2^2}{2 m_{\tilde q_1} M_h} \left( \frac{M_h T}{2\pi} \right)^{{3}/{2}} \!\!\!e^{-{M_h}/{T}}\!,
\end{eqnarray}
where low temperature expansion~\cite{Anderson:1991zb} has been used, and $M_h=126\,$GeV. 
Through Eq.~(\ref{Fx}), this causes the CP violation parameter $\varepsilon$ to vary with the temperature. 
Enough baryon asymmetry requires the source term for baryogenesis to remain effective,
i.e., $\varepsilon(T)>10^{-2}$, for a long enough period around the freeze out temperature $T_f\sim 20\,$GeV.
As a result, there is an upper bound on the differences of Yukawa couplings $|y_1^2 - y_2^2| <10^{-5}$.
This limits the choice of flavors to the nearly-degenerate sdown-sstrange co-LSP scenario only, which is able to give sufficient baryon asymmetry.
The right panel of Fig.~\ref{VaryLambda} shows an example of the evolution of $Y_B$ taking into account of the temperature dependence in $\varepsilon$.
In this case, for naturalness, the light stop can still be hidden by either cascade decaying to the LSPs or via its own RPV couplings.

\SEC{Summary. } To summarize, we propose a TeV scale baryogenesis picture, which can be realized in natural SUSY models, and can be directly probed in the LHC.
Baryogenesis from colored particle decays require CP violation to be close to maximal, which can be realized with nearly degenerate squarks.
The smoking-gun signal would be the charge asymmetry from the decay of squark resonances.
The charged lepton plus multijet final states without significant missing energy, should be taken into account in future analysis of LHC data, which will help to unveil the origin of baryon asymmetry in our universe.

\smallskip
\noindent{\bf Acknowledgement. } We acknowledge insightful discussions with Clifford Cheung, Tao Liu, Michael Ramsey-Musolf and Natalia Toro.
H.A.'s research at Perimeter Institute is supported by the Government of Canada through Industry
Canada and by the Province of Ontario through the Ministry of Research and Innovation.
Y.Z.'s work is supported by the Gordon and Betty Moore Foundation through Grant \#776 to the Caltech Moore Center
for Theoretical Cosmology and Physics, and by the DOE Grant DE-FG02-92ER40701.

\end{document}